\documentclass[twocolumn,showpacs,preprintnumbers,amsmath,amssymb,aps,prd,nofootinbib,10pt]{revtex4-1}
\usepackage{graphicx}
\usepackage{color}
\usepackage[utf8]{inputenc}
\usepackage[colorlinks,hyperindex]{hyperref}  
\usepackage{accents}
\usepackage{enumitem}
\hypersetup
 {
   colorlinks,%
   citecolor=blue,%
   linkcolor=blue,%
   urlcolor=blue,%
 }
 
\def\setR{\mathbb{R}}

\newcommand{\modiJu}[1]{{\color{black}{#1}}}

\begin{document}

\title{Massive scalar field on (A)dS space from a massless conformal field in $\setR^6$}

\author{E.~Huguet$^1$}
\author{J.~Queva$^2$}
\author{J.~Renaud$^3$}
\affiliation{$1$ - Universit\'e Paris Diderot-Paris 7, APC-Astroparticule et Cosmologie (UMR-CNRS 7164), 
Batiment Condorcet, 10 rue Alice Domon et L\'eonie Duquet, F-75205 Paris Cedex 13, France.\\
$2$ - Universit\'e de Corse -- CNRS UMR 6134 SPE,
    Campus Grimaldi BP 52, 20250 Corte, France.\\
$3$ - Universit\'e Paris-Est, APC-Astroparticule et Cosmologie (UMR-CNRS 7164), 
Batiment Condorcet, 10 rue Alice Domon et L\'eonie Duquet, F-75205 Paris Cedex 13, France.
} 
\email{huguet@apc.univ-paris7.fr, \\queva@univ-corse.fr,\\jacques.renaud@apc.univ-paris7.fr}

\date{\today}


\begin{abstract}
We show how the equations for the scalar field (including the massive, massless, minimally and conformally coupled cases) 
on de~Sitter and Anti-de~Sitter spaces can be obtained from both the SO$(2,4)$-invariant equation $\square \phi = 0$ in  $\setR^6$ 
and two geometrical constraints defining the (A)dS space. 
Thus, apart from the equation in $\setR^6$, the results  only follow from the geometry. 

\end{abstract}

\maketitle

\section{Introduction}\label{Sec-Introduction}

The relation between SO$(2,4)$-invariant free fields in $\setR^6$ and
fields in Minkowski space has long
been known \cite{Dirac:1936fq}. Nowadays, the space $\setR^6$ together with the conformal symmetry, SO$(2,4)$-invariance and/or 
Weyl rescaling invariance, provides a tool in working properties of (conformal) fields in four dimensional 
spaces \cite{Weinberg:2010fx, Weinberg:2012mz}. 
Besides  its usefulness, the existence of conformal symmetry as a (spontaneously broken) symmetry of Nature 
remains an open question  (see for instance \cite{Faria:2019lja} and references herein). This point provides 
a strong motivation for investigations on conformal symmetry.

In this paper, we show that scalar field equations on four dimensional de~Sitter (dS) and Anti-de~Sitter (AdS) spaces - whose solutions belongs to 
well known representations of the dS and AdS groups - can be obtained through a set of geometrical constraints acting on a massless SO$_0(2,4)$  scalar field in $\setR^6$. 
These constraints correspond to that used to restrict the phase space of a free particle of $\setR^6$
to the phase space of (A)dS. They are inherited from the two equations defining the (A)dS space as the intersection of  the null-cone of $\setR^6$ and a
hyperplane. For the free particle, these two equations supplement the hamiltonian constraint to reduce the phase space. For the \modiJu{classical} scalar field, they naturally define operators which can be diagonalized simultaneously with the Laplace-Beltrami operator
leading to the restriction of $\square \phi = 0$ to the (A)dS space. \modiJu{To satisfy the constraints, the solutions of $\square \phi = 0$ have to be homogeneous functions in $\setR^6$ of some unspecified degree of homogeneity. The restricted scalar equation on A(dS) 
is found to explicitly depends on this  degree of homogeneity. 
It can be freely set to recover the Klein Gordon equation on (A)dS whose coupling and mass parameters are known to label representations 
of the (A)dS group.}
A remarkable point is that the results come purely from the geometry: once the constraints defining the space (dS or AdS) as a sub-manifold 
in $\setR^6$ are given, the whole properties of the field, including its homogeneity, follow. Besides, the restricted scalar equation on
(A)dS can be compared to that appearing in \cite{Cardoso:2017qmj,Cardoso:2017egd}, allowing us to relate the operators introduced in that work 
to the degree of homogeneity of the field.

The paper is organized as follows. The geometrical framework is summarized in 
Sec. \ref{Sec-Geom}. We remind in particular how
the (A)dS space can be obtained from two geometrical constraints in $\setR^6$. 
We then briefly examine in Sec. \ref{Sec-FreePart} how these two constraints affect the free 
massless particle in $\setR^6$. This allows us to make apparent in Sec. 
\ref{Sec-KG} that the restriction of
the Laplace-Beltrami operator of $\setR^6$ to that of the (A)dS space comes, 
in our formalism, from purely 
geometrical constraints. We then obtain from a \modiJu{classical} massless scalar field \modiJu{equation} in $\setR^6$ 
the free Klein-Gordon scalar field on 
both de~Sitter and Anti-de~Sitter spaces. 
A proper choice of the degree of homogeneity, allows us to recover 
\modiJu{the equations which correspond to}
the well known representations of SO$_0(1,4)$ and SO$_0(2,3)$
-- including in particular the massive one -- and also  
equations obtained in \cite{Cardoso:2017qmj}.
\modiJu{We show that the set of solutions satisfying all 
the constraints contains solutions other than the conformal scalar,}
and then consider the action of the conformal group SO$_0(2,4)$.

We finally conclude. A reminder of the scalar 
representations of both de~Sitter and Anti-de~Sitter groups are given for reference in appendix \ref{Sec-ds-ads rep}. The link with the (A)dS equations examined in \cite{Cardoso:2017qmj} is provided in appendix \ref{Sec-MassLadderOperators}.
We briefly describe in appendix \ref{Sec-Interaction}  how a self-interaction term for the (A)dS field can be obtained pulled-back to $\setR^6$.


Our conventions are those of Fecko \cite{Fecko:2006} in differential geometry, for indexes they are: 
$\alpha, \beta, \gamma, \delta$, $\ldots = 0,\ldots, 5$, and $\mu,\nu,\rho,\sigma\ldots =  0, 1, 2, 3$.
Except otherwise stated, a superscript (or subscript) $f$ indicates a quantity on (A)dS space.

\section{Geometry}\label{Sec-Geom}

The space $\setR^6$ is provided with the natural orthogonal coordinates $\{y^\alpha\}$ and the metric  
$\eta~=~\mbox{diag}(+,-,-,-,-,+)$ which is left invariant under the group O$(2,4)$ and its connected subgroup SO$_0(2,4)$.

A four dimensional  manifold $X_f$ is obtained as the intersection 
of the five dimensional null 
cone of $\setR^6$ $\mathcal{C}:=\{y\in \setR^6 : C(y) = 0\}$,  where $C(y):=y^\alpha y_\alpha$, 
and a surface  $P_f:= \{y\in \setR^6 : f(y) = 1\}$, in which $f$ 
is homogeneous of degree one. 
Conformally flat spaces, including Robertson-Walker spaces,
 can be obtained by suitable choices of the 
homogeneous function $f$. Details about this construction may be found in \cite{Huguet:2013tv}.
Here we focus on dS and AdS spaces. The de~Sitter space is defined through
\begin{equation}\label{Eq-fdSitter}
f(y) := \frac{1}{2}(1+H^2)y^5 + \frac{1}{2}(1-H^2)y^4, 
\end{equation}
$H$ being related to the Ricci scalar through $R=-12 H^2$. 
The Anti-de~Sitter space is obtained by changing the parameter $H^2$ to $-H^2$ in the above equation. 

\modiJu{We note that the function (\ref{Eq-fdSitter}) is not the only possibility to obtain (A)dS spaces. Indeed, that choice of the function $f$ as already proven to be useful in the study of conformal fields.
In particular the flat space limit, from (A)dS to Minkowski space ($H=0$), becomes straightforward using this form  \cite{Huguet:2006fe}.
Since the equation for that field will appears as a special 
case in our study we keep the form (\ref{Eq-fdSitter}) for 
convenience.} 

\emph{In the remaining of the paper, except otherwise stated, we consider the de~Sitter space. The corresponding equations
in the Anti-de~Sitter space are obtained using the replacement} $H \rightarrow i H$.

To summarize, the definition of the $X_f$ manifold, here the (A)dS space, stems from the application of the two 
geometric constraints $C= 0$ and $f=1$.

\section{Constraints for the free massless particle}\label{Sec-FreePart}

In order to clarify the role of the constraints when we restrict a system from $\setR^6$
to the de~Sitter manifold $X_f$, let us consider the simple example of the ``free particle'' in $\setR^6$.
It satisfies the hamiltonian constraint: $H_6 = 0$,
where $H_6:=p^\alpha p_\alpha$ is the natural Hamiltonian in the phase space of $\setR^6$, namely T$^*\setR^6$. This space is
provided with its canonical one-form:\linebreak $\theta := p_\alpha dy^\alpha$.

The restriction to the de~Sitter space is expected to come from the constraints defining the de~Sitter manifold: 
$C = 0$ and $f=1$.
We observe that the phase space of $\setR^6$ is of dimension~$12$ and that of the 
de~Sitter space is of dimension~$8$. 
Thus, four constraints are needed in $\setR^6$. 
The two constraints which supplement the constraints $C$ and  $f$ 
are obtained from the Poisson brackets:
\begin{equation*}
\{H_6, C\} = 4 y^\alpha p_\alpha,~~\{H_6, f\} = 2  p_\alpha\partial^\alpha f,~~\{C, f\} = 0. 
\end{equation*}
The constraints $ D:=y^\alpha p_\alpha$ and $F:=(\partial^\alpha f)p_\alpha$
are secondary constraints in Bergmann-Dirac's terminology. Adding these 
constraints to the primary constraints $f, C$ and to the Hamiltonian leads to a closed Poisson algebra. 
Note that these 
constraints are, like $f$ and $C$,  geometrical in nature: they rewrite respectively as 
$D=\frac{1}{2}(\theta, dC)_\eta$ and  $F=(\theta,df)_\eta$ where $(\alpha,\beta)_\eta$ is the natural 
product between p-forms 
defined by $(\alpha,\beta)_\eta\mbox{vol}_\eta=\alpha\wedge\ast_\eta\beta$.

\section{The classical free scalar field}\label{Sec-KG}
We now consider the free Klein-Gordon equation in $\setR^6$, namely $\square_6 \phi = 0$, in which 
$\square_6$ is the Laplace-Beltrami operator. As in the previous section we look at the restriction of this equation to the  
space $X_f$.
\modiJu{We will see that on (A)dS the scalar fields fulfilling either the massless or massive Klein-Gordon equation
can be recovered solely from the geometrical constraints defined above.}

\modiJu{However, before addressing such a task we emphasize that our interest is focused on the interplay between the constraints expressed in $\mathbb{R}^6$ and the field equation to be satisfied.
The quantization of such fields from $\mathbb{R}^6$ is another program which we do not intend to undertake here.
Indeed, it would require to find a basis of mode solutions, equip them with a scalar product to build a proper Hilbert space and then study how this 
structure reduces on $X_f$.
Such a study is out of the scope of the present article and the latter point, which is by no means obvious in itself, is expected to be troublesome once cast on AdS (see the ``classical'' Ref.~\cite{Avis:1977yn}, or Ref.~\cite{Bertola:2000mx} for a different viewpoint).}

\subsection{Equations and constraints}\label{Sec-ConstEq}
As for the free particle we need some constraints in addition to $f$ and $C$ to perform the restriction 
of $\square_6 \phi = 0$ to $X_f$. 
One can expect that due 
to their geometrical nature the two constraints $D$ and $F$  will play a part in the restriction process. 
Indeed, the maps $p\mapsto D$ and 
$p\mapsto F$ define linear forms  on the phase space T$^*\setR^6$, they belong to 
T$^{**}\setR^6$, and thus they naturally identify with vectors of T$\setR^6$, namely $\frac{1}{2}\sharp dC$ and $\sharp df$ 
respectively. This justifies and replaces the usually postulated rule\footnote{Note that the 
Laplace-Beltrami operator $\square_6=\partial^\alpha\partial_\alpha$ can be obtained in 
the same way from the Hamiltonian $H_6=p^\alpha p_\alpha$.} $p_\alpha \mapsto \partial_\alpha$. 
We denote these vectors with the same symbols as before for simplicity: 
$D:= \frac{1}{2}\sharp dC = y^\alpha\partial_\alpha$ and $F := \sharp df =(\partial^\alpha f)\partial_\alpha$.

The two additional constraints $F$ and $D$ now are realized as operators.  Consequently, the restriction of 
$\square_6$ to $X_f$ appears as a particular case of the problem of finding the set 
of functions on which the operators $\square_6$, $D$, and $F$ are constants, that is to diagonalize them simultaneously. 
Then, the 
commutators 
\begin{equation*}
[\square_6, D] = 2 \square_6,~~~[\square_6, F] =  0,~~~[F, D] =  F,
\end{equation*}
show us that the only possibility is $\square_6 \phi= F\phi = 0$, $D\phi$ being unspecified, or more precisely that 
the function $\phi$ is homogeneous of some unspecified weight $r$, namely $D\phi = r \phi$. Note that the homogeneity 
of the field is not assumed, it is a consequence of the geometry.
To summarize, we will verify that to restrict $\square_6\phi = 0$ to the space $X_f$, corresponds to satisfy in $\setR^6$ the system
\begin{equation}\label{EQ-SystInR6}
\left \{
\begin{aligned}
\square_6 \phi&=0\\
F\phi &= 0 \\
D\phi &= r \phi.\\
\end{aligned}
\right .
\end{equation}

Now, the explicit expression of the restriction of $\square_6$ to the de~Sitter manifold is obtained  
by a straightforward calculation as follows.  We express $\square_6$, $D$ and $F$ in
the coordinate system (see for instance \cite{Huguet:2013tv})

\begin{equation*}
  x^c = \frac{y^\alpha y_\alpha}{(y^5+y^4)^2},\;
  x^\mu =  \frac{2y^\mu}{y^5+y^4},\;
  x^+ = f(y),
\end{equation*}
in which the two constraints $C$ and $f$, defining the de~Sitter space $X_f$, 
reads respectively $x^c=0$ and $x^+=1$. The dilation operator $D$ reads $x^+\partial_+$ and can be used to make apparent the 
degree of homogeneity $r$ of the scalar field $\phi$. After some algebra, the expression of $\square_6$ becomes
\begin{equation}\label{EQ-restriction}
\left(\square_{6}\phi\right)^f
= \square_f \phi^f   
+2(r+1)\left(F\phi\right)^f - r(r+3)H^2\phi^f,
\end{equation}
where $\square_f$ is the Laplace-Beltrami operator of de~Sitter space and $(A)^f$ is the restriction to $X_f$ 
of the scalar function $A$. 
We first note, that this expression does not depend on coordinates, as it is written in terms of geometrical objects, 
it is therefore a consequence of the geometry. 
Following the discussion of the previous paragraph $F\phi$ is now set to zero while
the homogeneity $r$ is left unspecified. 
Finally, we obtain the equation on the de~Sitter space $X_f$
\begin{equation}\label{EQ-EqdS}
\square_f \phi^f   - r(r+3)H^2\phi^f = 0.
\end{equation}

It is now obvious that this equation yields the free Klein-Gordon equation with a suitable choice of $r$. 
Besides that, one realizes that 
as far as the Minkowski spacetime, for which $H^2 = 0$, is considered only the massless
conformally invariant field fulfilling $\square_f\phi^f = 0$ can be
recovered.
On (A)dS spaces, however, thanks to a nonvanishing curvature a wider
class of \modiJu{equations} can be reached, encompassing \modiJu{those of} massive non-conformally
invariant fields. Hereafter, such cases are inspected on (A)dS space and are linked to the relevant 
unitary irreducible representations (UIRs) of group of isometries.

The homogeneity degree can also be related to the definition of the mass ladder operators introduced in \cite{Cardoso:2017qmj}. 
We make explicit the correspondence in appendix \ref{Sec-MassLadderOperators}.

Finally, let us note that a self-interaction term as the usual $\lambda\phi^4$ can be taken into account 
in the above derivation of the scalar equation (\ref{EQ-EqdS}). Nevertheless, it does not come from the 
geometrical constraints used here, we thus do not consider it further. We indicate for completeness how it can 
be obtained in appendix \ref{Sec-Interaction}. 

\subsection{\modiJu{Relating homogeneity to curvature and mass parameter}}\label{Sec-ConstRep}

The freedom left in the degree of homogeneity $r$ can be used to obtain various couplings. 
Since we are in particular interested in the massive  scalar let us set the parametrization 
\begin{equation*}
    - r(r+3)H^2\phi^f = (m^2 + 12 \xi H^2)\phi^f, 
\end{equation*}
$m$ and $\xi$ being parameters which, in the massive representations of SO$_0(1,4)$,
respectively identify with the mass of the field and the coupling constant to the scalar curvature $R = -12 H^2$. 
The above equation can be solved for $r$ and 
the symmetry $r\rightarrow -(r+3)$ allows us to choose the positive root without loss of generality 
\begin{equation*}
    r = -\frac{3}{2} + 
    \sqrt{\frac{9}{4} - \left(12 \xi +\frac{m^2}{H^2}\right)}.
\end{equation*}
This degree of homogeneity $r$, or more exactly the value of $12 \xi + (m/H)^2$, can be used in order 
to classify the scalar  UIRs of the de~Sitter group, they are reminded in appendix \ref{Sec-ds-ads rep}.

The case of the AdS space is obtained through the replacement $H \rightarrow i H$ in the de~Sitter formulas. The equation (\ref{EQ-EqdS}) 
becomes
\begin{equation}\label{EQ-EqAdS}
\square_f \phi^f   + r(r+3)H^2\phi^f = 0,
\end{equation}
and the corresponding reparametrization in terms of the ``mass'' $m$ and the coupling constant $\xi$ now reads: 
\begin{equation*}
    r(r+3)H^2\phi^f = (m^2 - 12 \xi H^2)\phi^f,
\end{equation*}
a solution of which is
\begin{equation*}
    r = -\frac{3}{2} +
    \sqrt{\frac{9}{4} - \left(12 \xi -\frac{m^2}{H^2}\right)},
\end{equation*}
where we kept the positive root only, since it gives the right homogeneity degree $r= -1$ for the massless conformally
coupled  scalar field.

The value of $12 \xi - (m/H)^2$  relates the field to scalar UIRs of SO$_0(2,3)$, the 
procedure is reminded in appendix~\ref{Sec-ds-ads rep}.

\subsection{Solutions for the non-conformal field}
We know already that the set of functions satisfying the system (\ref{EQ-SystInR6})
is non-empty: it contains at least the conformal scalar for which $r=-1$ \cite{Huguet:2006fe}.
In this section we note that \modiJu{solutions} for other values of $r$ can also be found
\modiJu{, for instance} the set of functions  $\phi_{k,r}(y):=(k\cdot y)^r$, 
where $k$ is a constant vector of $\setR^6$, satisfy
\begin{align*}
\square_6 \phi_{k,r} &= r(r-1) k^2  \phi_{k,r-2},\\
F\phi_{k,r} &= r f( k) \phi_{k,r-1},\\
D\phi_{k,r} &= r \phi_{k,r},
\end{align*}
from which it is apparent that $\{\phi_{k,r}(y)\}$ with $k^2=0$ and $f(k) = [(1+H^2)k^5 +(1-H^2)k^4]/2= 0$, is a set of 
\modiJu{solutions}  for Eq. (\ref{EQ-EqdS}) 
on de~Sitter space, that is for $y\in X_f$.

The study of these \modiJu{solutions} is out of the scope of the present paper. We just remark that when $(k\cdot y)$ vanishes these 
\modiJu{solutions} are ill-defined. \modiJu{Nevertheless, due to the above constraints, the vanishing of $(k \cdot y)$ only occurs on a submanifold 
of $X_f$. At least on de~Sitter space
the problem can be circumvented 
following the} method used in 
\cite{Bros:1994dn}, where the \modiJu{solutions} are seen as boundary values of their complexified versions in a suitable domain 
of analyticity. \modiJu{Then, there exist solutions other than the conformaly coupled (massless) scalar.}

\subsection{Group action}\label{Sec-Group}
Since the constraint $F\phi = 0$  refers explicitly to the space $X_f$ we expect the invariance of the set of 
solutions of system (\ref{EQ-SystInR6}) to be restricted to the  (A)dS group. This is indeed the case, with the 
known exception of the conformal scalar field. 

Let us recall the algebra of so$(2,4)$ 
\begin{equation*}
\left[X_{\alpha \beta}, X_{\gamma \delta}\right] =  \eta_{\beta \gamma} X_{\alpha \delta} +\eta_{\alpha \delta} X_{\beta \gamma} 
- \eta_{\alpha \gamma} X_{\beta \delta} - \eta_{\beta \delta} X_{\alpha \gamma}, 
\end{equation*} 
which realizes on functions with  $X_{\alpha \beta}~=~y_\alpha \partial_\beta -  y_\beta \partial_\alpha$. One can check 
that these generators commute 
with both $\square_6$ and $D$. By contrast, they do not commute with the operator $F$ except for
$X_{\mu\nu}$ and the combination 
\begin{equation*}
{\displaystyle Y_\mu := \frac{1}{2} (1 - H^2) X_{5 \mu} - \frac{1}{2} (1 + H^2) X_{4 \mu}},
\end{equation*} 
which together form a (A)dS sub-algebra. 
We thus conclude that the set of solutions of the system (\ref{EQ-SystInR6}) is at least invariant under the (A)dS group. 

For the special case of the conformal scalar, whose degree of homogeneity is $r=-1$, the equation (\ref{EQ-restriction}) 
shows that the effect of the constraint $F$ disappears. Consequently, the field remains invariant under the whole conformal group 
SO$_0(2,4)$.
The explicit group action in that case includes a multiplicative conformal weight
$\omega^f_g$ which accounts for the group elements which are not in the
(A)dS group.
This point as already been described in our geometrical framework in \cite{Huguet:2008js, Huguet:2013tv}.
\newline

\section{Conclusion}
In this work we have shown that the equations for the scalar fields 
\modiJu{(massless, massive, conformally invariant or not, \ldots) on (A)dS space} are obtained from the massless scalar equation $\square_6 \phi = 0$ in $\setR^6$ 
(endowed with the metric $\eta = \mathrm{diag}(+,-,-,-,-,+)$) from  purely geometrical considerations. Precisely, the constraints defining 
the (A)dS spaces from $\setR^6$ are enough to reduce the Laplace-Beltrami operator $\square_6$ to that of (A)dS spaces with the condition that the scalar field be homogeneneous of some unspecified degree $r$. The resulting 
equation on (A)dS is a Klein-Gordon equation with a mass-term parametrized by $r$. This parameter can be choosen in order to recover the UIRs of the (A)dS group. Besides, it can be related to the mass ladder operators defined in \cite{Cardoso:2017qmj}.

We also notice that a self-interacting term 
can be pulled back from (A)dS to $\setR^6$. Since the formalism we use allows us to obtain 
solutions of free scalar equations on (A)dS from solutions of $\square_6 \phi = 0$ in $\setR^6$,
it could be interesting to investigate if an interacting theory in $\setR^6$ can be built in order to obtain solutions in
the curved (A)dS space-times? 
\appendix
\section{Scalar representations of (A)dS group}\label{Sec-ds-ads rep}
\modiJu{On de~Sitter space, the Laplace-Beltrami operator $\square_f$,  and the first order Casimir operator 
$\mathcal{Q}_1$,  are related through $\square_f = - H^2 \mathcal{Q}_1$, then solutions are related to scalar UIRs.}
For the de~Sitter SO$_0(1,4)$ the scalar irreducible representations are labeled as (see for instance \cite{Gazeau:2008zz}):
\begin{itemize}
\item ${\displaystyle 12 \xi +\frac{m^2}{H^2}\geqslant \frac{9}{4}}$: Principal series of representations.
\item ${\displaystyle 0<12 \xi +\frac{m^2}{H^2}<\frac{9}{4}}$: Complementary series of representations with 
the special case of the massless conformally coupled field obtained for $\xi= 1/6$, $m=0$ and $r=-1$.
\item ${\displaystyle 12 \xi +\frac{m^2}{H^2} = 0}$: The massless minimally coupled field, that is the first term 
of the discrete series of representations \cite{Gazeau:1999mi}. 
\end{itemize}

The case of the scalar irreducible representations of the AdS SO$_0(2,3)$ group is performed as follows.

On the one hand, the UIRs of SO$_0(2,3)$ are labelled as $\mathcal{D}(E_0,s)$, see
\cite{Angelopoulos:1980wg,Gazeau:2008zz}.
In the scalar case, here considered, one has
$\mathcal{D}\left(E_0 \geqslant 1/2, 0\right)$ for which the spectrum of
the first order Casimir operator $\mathcal{Q}_1$ reduces to
$\mathcal{Q}_1 = E_0(E_0 -3)$.
The spectrum of $\mathcal{Q}_1$ is consequently bounded from below with
$\mathcal{Q}_1 \geqslant - 9/4$, since $E_0 \geqslant 1/2$.

On the other hand, on AdS in the scalar representation one can
relate $\mathcal{Q}_1$ to the wave equation through
$\square_f =  H^2 \mathcal{Q}_1$. Then, the above bound translates into the condition
\begin{equation*}
    - \frac{9}{4} \leqslant \left(12 \xi - \frac{m^2}{H^2}\right).
\end{equation*}
This condition is in direct relation to the Breitenlohner-Freedman bound 
\cite{Breitenlohner:1982jf}, for which a conserved convergent scalar product
might be defined upon the space of solutions to the wave equation.

Thus, provided the above condition is fulfilled the relevant UIRs can be retrieved:

\begin{itemize}
\item 
${\displaystyle   - \frac{9}{4} < 12 \xi - \frac{m^2}{H^2} \leqslant - \frac{5}{4}}$:\\Two UIRs might be involved, 
namely $\mathcal{D}\left(3/2 \pm \sqrt{9/4 + 12 \xi - (m/H)^2}, 0\right)$.
Those UIRs can be distinguished by their lowest eigenvalue of the
``conformal energy'' $E_0$. 
In this specific range lies the massless conformally coupled field, which 
belongs to the reducible representation $\mathcal{D}(1,0)\oplus \mathcal{D}(2,0)$, obtained for $\xi= 1/6$, $m=0$ and $r=-1$. 
At the edge of this interval, for
$12 \xi - (m/H)^2 = - 5/4$,
lies the, so-called, $\mathcal{D}(1/2, 0)=\mathrm{Rac}$ representation
\cite{Flato:1980we}.
In this interval both the ``regular'' and ``irregular'' modes are allowed, see
\cite{Breitenlohner:1982jf}. 

\item 
${\displaystyle -\frac{5}{4} < 12 \xi - \frac{m^2}{H^2}} $: Only one UIR
might be involved, namely: $\mathcal{D}\left(3/2 + \sqrt{9/4 + 12 \xi - (m/H)^2}, 0\right)$.
There only the ``regular'' modes are allowed.

\end{itemize}
%
%
\section{Relation with mass-ladder-operators formalism }\label{Sec-MassLadderOperators}
In \cite{Cardoso:2017qmj} Cardoso, Houri and Kimura, introduce mass-ladder-operators $D_k$, $k$ being a parameter, 
which map the scalar field solution of the Klein-Gordon equation $(\square - M^2) \Phi = 0$ to the equation
$(\square - (M^2 +\delta M^2)) D_k\Phi = 0$. For (A)dS in four dimensions, using notation of \cite{Cardoso:2017qmj}, 
one has $M^2 = -\chi k(k+3)$, which correspond to our Eq. (\ref{EQ-restriction}). Thus in our notation $\chi = \pm H^2$ 
and $k = r$. In order to exist, the operators $D_k$ are subject to the conditions (Eqs. (9) of \cite{Cardoso:2017qmj}):
\begin{equation*}
\left\{
    \begin{aligned}
    \displaystyle\frac{9}{4} &\chi \leqslant M^2,~\chi<0~(AdS),\\
    \displaystyle\frac{9}{4} &\chi \geqslant M^2,~\chi>0~(dS). 
    \end{aligned}
    \right.
\end{equation*}
They correspond to UIRs of (A)dS for fields uncoupled to the background space~: $\xi = 0$ as seen on expressions reminded in 
appendix \ref{Sec-ds-ads rep}. Two kind of operators  can be defined (Eqs. (10) of \cite{Cardoso:2017qmj}) they are labelled by
\begin{equation*}
k_\pm = -\frac{3}{2} \pm
    \sqrt{\frac{9}{4} -\frac{M^2}{\chi}}.
\end{equation*}
From Sec. \ref{Sec-ConstRep}, it appears that for $H^2 > 0$ (dS), both roots ($k_\pm$) coincide with the degree of homogeneity $r$ of the (uncoupled) field; for $-H^2 < 0$ (AdS), only the positive root ($k_+$) corresponds to the degree of homogeneity.

%
%
\section{About self-interaction}\label{Sec-Interaction}
The homogeneity of both the scalar field and the function $f$ gives us the possibility to obtain 
the scalar equation (\ref{EQ-EqdS}) with an interaction term
\begin{equation}\label{EQ-ScalInt}
\square_f \phi^f   - r(r+3)H^2\phi^f = \lambda \left(\phi^f\right)^n, 
\end{equation}
from the corresponding equation in $\setR^6$. 
In effect, let us first recall that since $f$ is homogeneous 
of degree one the 
point $y^f := y/f(y)$
belongs to the plane $P_f$ whose intersection with the null cone is $X_f$. This can be seen on the equalities: 
$f(y^f) = (f(y))^{-1}f(y)=1$.   
In particular, the term $\lambda f^s(y) \phi^n(y)$, for some powers $s$ and $n$ and some constant $\lambda$, 
restricted to $X_f$ is: 
\begin{equation*}
\lambda f^s(y^f) \phi^n(y^f) = \lambda \left(\phi^f\right)^n(y^f).
\end{equation*}
Now, for a given $n$, let us choose $s$  such that 
$r-2= s + rn$. Then, the equation
\begin{equation*}
\square_6 \phi = \lambda f^s \phi^n,
\end{equation*}
is meaningful regarding the homogeneity. Now,
when it is restricted to $X_f$ it becomes the scalar equation (\ref{EQ-ScalInt}).
%
%
\bibliography{KGMassive-HugQueRenV3-bibliography}
\end{document}